\font\tx=cmr10 at 11pt  \font\ma=cmmi10 at 11pt  \font\sy=cmsy10 at 11pt
\textfont0=\tx          \textfont1=\ma           \textfont2=\sy
\font\sub=cmr8          \font\masub=cmmi8        \font\sysub=cmsy8
\scriptfont0=\sub       \scriptfont1=\masub      \scriptfont2=\sysub
\font\title=cmbx12      \font\author=cmr12       \font\bx=cmbx10 at 11pt
\def\space{\vskip 12.5pt}
\def\section#1{\space\space\goodbreak\leftline{\bx#1}\nobreak\space}
\def\subsection#1{\space\goodbreak\leftline{\bx#1}\nobreak\space}
\def\Msun{M$_\odot$}    \def\M/L{$M\!/\!L$}
\def\m1{$m_1$}          \def\mp{$m_{\rm p}$}
\baselineskip=12.5pt    \parindent=18pt
\raggedbottom           \tx
\vglue 1cm
\centerline{\title  EARLY STAR FORMATION AND THE EVOLUTION OF THE}
\medskip
\centerline{\title    STELLAR INITIAL MASS FUNCTION IN GALAXIES}
\space\space
\centerline{\author              Richard B. Larson}
\space
\centerline{      Yale Astronomy Department, Box 208101, New Haven,
                                 CT 06520-8101, USA}
\centerline{                   larson@astro.yale.edu}
\space
{\narrower
\section{ABSTRACT}

    It has frequently been suggested in the literature that the stellar
IMF in galaxies was top-heavy at early times.  This would be plausible
physically if the IMF depends on a mass scale such as the Jeans mass
that was higher at earlier times because of the generally higher
temperatures that were present then.  In this paper it is suggested, on
the basis of current evidence and theory, that the IMF has a universal
Salpeter-like form at the upper end but flattens below a characteristic
stellar mass that may vary with time.  Much of the evidence that has
been attributed to a top-heavy early IMF, including the ubiquitous
G-dwarf problem, the high abundance of heavy elements in clusters of
galaxies, and the high rate of formation of massive stars in
high-redshift galaxies, can be accounted for with such an IMF if the
characteristic stellar mass was several times higher during the early
stages of galaxy evolution.  However, significant variations in the
mass-to-light ratios of galaxies and large amounts of dark matter in
stellar remnants are not as easily explained in this way because they
require more extreme and less plausible assumptions about the form and
variability of the IMF.  Metal-free `population~III' stars are predicted
to have an IMF that consists exclusively of massive stars, and they
could help to account for some of the evidence that has been attributed
to a top-heavy early IMF, as well as contributing importantly to the
energetics and chemical enrichment of the early universe.
\space}

\section{1~~INTRODUCTION}

\noindent
    The stellar initial mass function (IMF), or distribution of masses
with which stars are formed, is of fundamental importance in determining
the properties of stellar systems and how they evolve with time; even
the processes of galaxy formation cannot be adequately understood
without a knowledge of the IMF at early times.  Efforts to model the
formation and evolution of galaxies have generally assumed that the IMF
is a universal function that does not vary with time, yet it has long
been recognized that many of the predictions of such models would be
altered importantly if the IMF were to vary.  This is because the
present luminosities of galaxies depend mostly on the number of stars
they contain with masses near one solar mass, but their heavy-element
contents and the energy feedback effects that control the evolution of
their gas contents depend on the number of stars that formed in them
with masses above ten solar masses; thus, if the relative numbers of
low-mass and high-mass stars produced had been different at early
times, such basic properties of galaxies as their luminosities, colors,
and chemical compositions and their evolution with time would be
substantially altered.  This would have far-reaching implications
for the interpretation of many observations of galaxies, especially
observations at high redshifts.

    For many years, beginning with Schwarzschild \& Spitzer (1953),
there have been speculations in the literature that the IMF was
dominated by massive stars at early times.  Some of the evidence for a
variable IMF, and some of the possible implications of models of galaxy
evolution with an IMF that was more top-heavy at early times, were
discussed earlier by Larson (1986a,b, hereafter L86).  Although it now
appears that the specific model proposed by L86 was probably too extreme
(Meusinger 1989; Larson 1992b), a number of lines of evidence have
continued to suggest that the IMF in galaxies may have been biased
toward massive stars at early times.  The following types of evidence
will be discussed in some detail in this paper, along with their
possible interpretation in terms of a time-varying IMF:

\medskip \noindent
(1) The original reason for proposing that the early IMF was dominated
by massive stars was that standard hot-big-bang cosmology predicts that
the first stars contained no heavy elements, yet no metal-free stars
have ever been found.  Although it is possible that the number of such
stars formed was negligible, this is not very plausible because a
non-negligible amount of time must have elapsed between the formation
of the first metal-free stars and the appearance of the first stars
polluted with their nucleosynthetic products.

\medskip \noindent
(2) A similar problem of long standing is the classical `G-dwarf
problem', namely the deficiency of metal-poor stars in the solar
neighborhood compared with the predictions of simple models of chemical
evolution.  A related fact is the observation that the abundances of
heavy elements in the Galactic disk have shown only a slow increase
with time over most of Galactic history.  These observations suggest
that the rate of production of heavy elements was higher at early times
than at more recent times.  

\medskip \noindent
(3) The large total mass of heavy elements observed in the hot gas in
large clusters of galaxies suggests that the early IMF in these systems
contained a higher proportion of massive stars than the present-day
IMF\null.  The increase of both the metallicities and the mass-to-light
ratios of cluster galaxies with mass also suggests that the IMF may have
been more top-heavy in the most massive galaxies, which could have been
the first to form their stars.

\medskip \noindent
(4) There is possible evidence that the gas in clusters of galaxies was
heated by more early supernovae than would have been predicted by a
standard IMF, and this again suggests a top-heavy early IMF in clusters.

\medskip \noindent
(5) Recent studies of galaxy populations at high redshifts have
suggested that a top-heavy early IMF may be needed to account for the
strong evolution of the cosmic luminosity density with redshift,
especially if there is significant dust extinction in high-redshift
galaxies.

\medskip \noindent
(6) Standard hot-big-bang cosmology suggests that most of the baryonic
matter in the universe is in a dark form, and one possibility could be
the dark remnants of early generations of massive stars.  The claimed
detection via microlensing of stellar-mass `machos' in the halo of our
Galaxy has led to the suggestion that a significant amount of dark
matter might be in stellar remnants.
\medskip

     Most of the models that have been developed to describe the
formation and evolution of galaxies have assumed that the stellar IMF
is a power law similar to that originally proposed by Salpeter (1955);
if any variability of the IMF has been allowed, this variability has
usually been assumed to take the form of a change in the slope of the
power law.  However, this assumption is not supported by recent evidence
suggesting that the IMF in many nearby systems always has a slope
similar to the Salpeter value above a solar mass, while it flattens at
lower masses, possibly even turning down below a few tenths of a solar
mass.  The departure of the IMF from a power law at the low end might
reflect the existence of a mass scale in the star formation process
that depends on cloud properties such as temperature and pressure, and
such a mass scale might be expected to vary with place and time.  In
this case, another possible form for a varying IMF might be given by a
function that has a universal power-law form at large masses but departs
from this power law below a characteristic mass that can vary.  The mass
scale for star formation probably depends most strongly on temperature,
and since the temperature in star-forming clouds was almost certainly
higher at earlier cosmic times, the mass scale should also have been
higher at earlier times.  This would have the effect of decreasing the
relative number of low-mass stars formed at early times.

    In this paper it will be shown that many of the observations
mentioned above can be accounted for with an IMF whose mass scale was
higher at early times, and the constraints that can be placed on such
models will be discussed.  Section~2 summarizes current evidence and
theory regarding the present-day IMF, and Section~3 discusses how the
associated mass scale might be expected to vary with time.  Section~4
considers whether such a time-varying IMF can account for the chemical
evidence mentioned in items (1)--(3) above and the thermal evidence
mentioned in item~(4).  Section~5 discusses the possible evidence for
variability of the IMF provided by high-redshift observations, and
Section~6 discusses the extent to which models with a top-heavy early
IMF might account for dark matter and machos.  Finally, Section~7
discusses the possibility that metal-free `population~III' stars with an
IMF strongly biased toward high masses might have existed in significant
numbers and might help to account for some of the observations mentioned
above.  The conclusions of the paper are summarized in Section~8.

\section{2~~EVIDENCE AND THEORY CONCERNING THE PRESENT-DAY IMF}

\noindent
     As has been known since the work of Salpeter (1955), the
present-day IMF of stars in the solar neighborhood can be approximated
by a declining power law for masses above a solar mass, but it falls
below an extrapolation of this power law at lower masses, flattening
below about 0.5$\,$\Msun\ and possibly even declining (in number of
stars per unit logarithmic mass interval) below 0.25$\,$\Msun\ (Scalo
1986, 1998).  The behavior of the IMF at the lowest masses remains very
uncertain because of the poorly known mass-luminosity relation for the
faintest stars; most studies have suggested that the IMF is roughly flat
(in logarithmic mass units) at the low end, but a significant falloff
below 0.1$\,$\Msun\ seems indicated by the relative paucity of brown
dwarfs (Basri \& Marcy 1997).  The IMF in open clusters is generally
similar to that of the field stars, although possibly more variable; the
best-studied young cluster, the Orion Nebula cluster, has an IMF very
similar to that derived by Scalo (1986) for field stars, including a
decline below 0.2$\,$\Msun\ (Hillenbrand 1997).  Thus, present evidence
allows the possibility that the IMF in the solar neighborhood may be
roughly flat at the low end, or that it may peak at a mass of around
0.25$\,$\Msun\ and decline into the brown-dwarf regime, or even that it
may be quite variable at the low end.  Whether or not there is a peak,
the flattening of the IMF below a solar mass implies that there is a
characteristic stellar mass of the order of one solar mass such that
half of the mass that condenses into stars goes into stars above this
mass and half goes into less massive stars.

    Recent studies of the IMF in clusters and associations in our Galaxy
and the Magellanic Clouds have generally supported the universality of
the power-law part of the IMF above 1$\,$\Msun, and have shown that in
most cases the slope $x$, defined such that $dN\! /d\log m \propto
m^{-x}$, is not very different from the original Salpeter value
$x = 1.35$ (von Hippel et al.\ 1996; Hunter et al.\ 1997; Massey 1998). 
However, Scalo (1998) emphasizes the large scatter that still exists in
these results, and notes possible trends with mass such that the slope
at intermediate masses may be significantly steeper than the Salpeter
value and equal to $x \sim 1.7 \pm 0.5$, similar to what was found
earlier by Scalo (1986).  Massey (1998) also finds evidence that the
slope is steeper for field stars than for clusters and associations. 
Nevertheless, clusters and associations probably yield more reliable
results than field stars because fewer assumptions are required to
derive their IMFs, and the Salpeter slope is representative of the
results found for them at masses above a solar mass, as is illustrated
in Figure~5 of Scalo (1998).  At lower masses the IMF slope clearly
becomes much smaller than the Salpeter value, and estimates of the slope
below 0.5 solar masses generally fall in the range $x \sim 0 \pm 0.5$. 
The IMF below 0.1$\,$\Msun\ remains very poorly known because only a
small number of objects have yet been found in this mass range, and
suggested values of the slope in the brown-dwarf regime range from
$x \sim -1$, implying an IMF that declines significantly at the low end,
to $x \sim 0$, implying an IMF that is flat at the low end (e.g., Basri
\& Marcy 1997; Mart\'\i n et al.\ 1998; Mayor, Queloz \& Udry 1998;
Festin 1998).

     Since the actual IMF thus remains quite uncertain, especially at
the low end, we shall consider two simple approximations that encompass
the range of possibilities suggested by the evidence mentioned above. 
If brown dwarfs are indeed rare objects and the IMF peaks at mass of a
few tenths of a solar mass, its shape may be similar to that of the
analytic function adopted by L86 to represent a low-mass mode of star
formation, $dN\! /d\log m \propto m^{-2}\exp[-(m_1/m)^{1.5}]$, except
for being less sharply peaked than this function and falling off more
gradually toward both lower and higher masses.  We therefore consider,
as one possibility, the following simple analytic form:

$$       dN\! /d\log m \propto m^{-1.35}\exp(-m_1/m).       \eqno(1)$$

\noindent This function has a logarithmic slope $x = 1.35 - m_1/m$,
so it approaches a power law with the Salpeter slope $x = 1.35$ at
large masses, peaks at a mass $m_{\rm p} = m_1/1.35$, and falls off
exponentially with increasingly negative $x$ at lower masses.  Since
this function has a steeper falloff at the low end than is suggested by
most of the evidence mentioned above, we consider also the possibility
that the IMF does not decline at all at the low end.  If brown dwarfs
are as common as is suggested by the most optimistic recent estimates,
and if the IMF accordingly is approximately flat at the low end, it may
be represented approximately by the following simple alternative form:

$$        dN\! /d\log m \propto (1 + m/m_1)^{-1.35}.        \eqno(2)$$

\noindent This function is very similar to (1) at masses above \m1\ and
has a logarithmic slope $x = 1.35(1 + m_1/m)^{-1}$, so that it again
approaches the Salpeter form at large masses but becomes asymptotically
flat with $x = 0$ at the low end.  Approximations (1) and (2) are thus
consistent with the evidence summarized above that the IMF typically has
a Salpeter form for large masses, while they allow for a large range of
uncertainty or variability of the lower IMF, and they also allow for the
possibility that the mass scale \m1\ may be variable.

    The mass scale \m1\ might be expected to be related to a fundamental
scale in the star formation process such as the Jeans mass, and evidence
supporting this possibility has been discussed by Larson (1995, 1996,
1998).  Analyses of the spatial clustering of the T~Tauri stars in the
Taurus and Ophiuchus regions have revealed the existence of two distinct
regimes, a hierarchical clustering regime on large scales and a binary
regime on small scales, with a break occurring at a separation of about
0.05$\,$pc in a plot of average companion surface density versus
separation (Larson 1995; Simon 1997).  The existence of this break
suggests that the clustering hierarchy is built up of basic star-forming
units with a radius of $\sim 0.05\,$pc that typically form binary stars.
These units can be identified with the `ammonia cores' in which T~Tauri
stars form, and these cores have masses of the order of 1$\,$\Msun,
consistent with the possibility that they typically form two stars with
masses of $\sim 0.5\,$\Msun.  These radii and masses are very similar to
the radius and mass of a marginally stable Bonnor-Ebert sphere with a
temperature of 10$\,$K and a boundary pressure equal to the typical
non-thermal pressure of $\sim 3 \times 10^5$ cm$^{-3}\,$K observed in
molecular clouds (Larson 1996, 1998).  The Jeans scale calculated in
this way can be regarded as the scale at which there is a transition
between a chaotic regime dominated by non-thermal pressures on larger
scales and a regular regime dominated by thermal pressure on smaller
scales; whatever the details of the star formation process, this
transition scale almost certainly plays some role in determining
stellar masses and the IMF.

    Direct evidence for a transition between different physical regimes
on different scales has been provided by studies of the internal
kinematics of star-forming cloud cores by Goodman et al.\ (1998).  On
scales larger than about 0.1$\,$pc, the non-thermal component of the
velocity dispersion in these cores increases with region size following
the general linewidth-size relation for molecular clouds, while on
smaller scales the non-thermal velocity dispersion becomes smaller than
the sound speed and almost independent of region size.  Goodman et al.\
interpret these results as indicating a transition from a regime of
chaotic motion on larger scales to a regime of `velocity coherence' on
smaller scales, and they note that this transition occurs at a scale
similar to that found for the break in the clustering properties of
T~Tauri stars mentioned above and suggest that both results reflect the
existence of an `inner scale' of a self-similar process, i.e.\ a scale
at which there is a transition from chaotic behavior on larger scales
to regular behavior on smaller scales.  If the physical basis of this
transition is the transition from dominance by non-thermal pressures on
large scales to dominance by thermal pressure on small scales, the
associated scale is just the Jeans scale as defined above.  Thus, both
theoretical and observational considerations suggest that there is an
intrinsic scale in the star formation process that can be identified
with the Jeans scale.  The power-law upper part of the IMF may be
produced by self-similar accumulation processes occurring in the chaotic
regime, possibly arising from the universal scale-free dynamics of
turbulence and possibly describable by fractal geometry (Larson 1992a,
1995; Elmegreen 1997).

    The Jeans mass depends on the temperature and pressure in
star-forming clouds and is proportional to $T^2P^{-1/2}$.  In
self-gravitating clouds the pressure is proportional to the square of
the surface density $\mu$, and the Jeans mass is therefore proportional
to $T^2\mu^{-1}$, a result also found from stability analyses of
self-gravitating flattened or filamentary structures (Larson 1985). 
Because of the strong dependence of this mass scale on the temperature,
it seems likely that the temperature in star-forming clouds is the most
important parameter controlling the mass scale for star formation
(Larson 1985), although the effect of changes in temperature can be
compensated by sufficiently large changes in pressure or surface
density.  The temperature is controlled by the balance between heating
by external radiation fields and cooling by line emission from atoms
and molecules and thermal emission from dust, while the pressure
is controlled by the dynamical processes that create and confine
star-forming clouds (Larson 1996).  More intense external radiation
fields and lower heavy-element abundances would both imply higher cloud
temperatures, and thus would increase the mass scale for star formation
unless there is a sufficiently large compensating increase in pressure. 
These effects provide a basis for order-of-magnitude estimates of how
the IMF might vary with time or metallicity in galaxies, as will be
discussed in the next section.  Note that a strong dependence of the
mass scale for star formation on cloud temperature is expected on quite
general grounds; for example, it is also predicted by the hypothesis
that stellar masses are controlled by a temperature-dependent accretion
rate (Adams \& Laughlin 1996).

\section{3~~ POSSIBLE TIME VARIABILITY OF THE IMF}

\noindent
    Past studies exploring the effect of a time-varying IMF have mostly
assumed either a power-law IMF with a variable slope (Schmidt 1963) or
a bimodal IMF with high-mass and low-mass components that can vary
independently (L86).  However, as has been noted above, present evidence
does not support either a variable slope or the existence of bimodality
or any other kind of `fine structure' in the IMF; in particular, the
kink near 1$\,$\Msun\ that was considered by Scalo (1986) and L86 to
provide possible evidence for bimodality now seems better interpreted as
an artifact resulting from an inaccurate mass-luminosity relation.  The
evidence and theory summarized above suggest instead that the IMF may
have a universal slope at the upper end, and that any variability may
be confined to the lower end, possibly being associated with a variable
mass scale \m1\ at which it departs from a power law.  The essential
effect of this type of variability is to alter the relative number of
low-mass stars formed, compared with the number of more massive stars;
if the mass scale was higher at earlier times, relatively fewer low-mass
stars would have been formed at these times.

    It was suggested above that the mass scale of the IMF depends mainly
on the temperature in star-forming clouds.  This temperature was almost
certainly higher at earlier times for several reasons: (1) The cosmic
background temperature, which is the minimum possible cloud temperature,
increases with redshift $z$ as $T = 2.73(1 + z)$, and it becomes higher
than the present minimum temperature of $\sim 8\,$K in molecular clouds
at redshifts greater than 1.9.  (2) The generally lower metallicity of
the gas at early times implies lower cooling rates, and hence higher
temperatures for a given heating rate.  (3) The heating rate was almost
certainly higher at early times because of the more intense radiation
fields that were present at all wavelengths; in particular, ultraviolet
radiation from young stars and cosmic rays from supernovae, which are
the most important heating effects at the present time, would both have
been more important at early times when the star formation rate per unit
volume was higher.  (4) Starburst activity, which may produce a locally
top-heavy IMF for similar reasons, may have been more prevalent during
the early stages of galactic evolution, especially if mergers played an
important role.

    Since the Jeans mass depends on the pressure as well as the
temperature of star-forming clouds, the effect of higher temperatures
could have been compensated by much higher pressures.  However, there is
no apparent reason why the pressures in star-forming clouds should have
been much higher at earlier times.  At the present time, the typical
pressure in star-forming molecular clouds is already much higher than
the general pressure of the interstellar medium, probably as a result of
the thermal and dynamical feedback effects of star formation on the ISM
(Larson 1996).  At early times, the feedback effects that pressurize
present-day molecular clouds might not yet have become established,
and such large pressure enhancements might not have been common. 
The extreme pressures needed to form low-mass stars at the higher
temperatures present at early times might have existed only in
exceptionally dense and massive clouds, such as those that may have
formed the globular clusters.  If so, the formation of low-mass stars
at early times might have occurred only in relatively extreme
circumstances, and globular clusters might not be representative of
early star formation.

    Thus, while it seems likely that the IMF was generally more
top-heavy at earlier times, many physical effects were probably involved
and no quantitative predictions can yet be made.  However, it is worth
noting that even the minimal effect of a higher cosmic background
temperature can significantly increase the Jeans mass during early
stages of galaxy evolution.  Even if the cloud temperature increases
by the minimum possible amount with redshift and is equal to either
the present 8$\,$K or the background temperature $2.73(1 + z)$ when
the latter is higher, the predicted Jeans mass for a given pressure
increases by a factor of 2 at a redshift of~3, and by a factor of 4
at a redshift of~5.  If the minimum cloud temperature is twice the
background temperature at the higher redshifts, the Jeans mass for
a given pressure increases by a factor of 7 at a redshift of~3 and a
factor of 17 at a redshift of~5.  Even though the possible compensating
effect of a higher pressure is difficult to estimate, these examples
suggest that the mass scale of the IMF could plausibly have been several
times higher during the early stages of galaxy evolution, and higher
by an order of magnitude or more at redshifts larger than~5.

    To illustrate the possible consequences of such an increase with
redshift in the mass scale of the IMF, if the peak mass \mp\ in IMF (1)
is increased by a factor of 4 from 0.25 to 1.0$\,$\Msun, the ratio of
the number of solar-mass stars formed to the number of very massive
stars formed is decreased by a factor of 2.8; for IMF (2), which is
flatter at the low end, this ratio is decreased by a factor of 2.1 for
the corresponding increase in the mass scale \m1.  If the mass scale of
the IMF is increased by a further factor of 2, the ratio of solar-mass
stars to very massive stars is decreased by a factor of 11 for IMF (1)
and a factor of 4 for IMF (2).  Such changes are large enough to have
very important consequences for galactic evolution, as will be discussed
further below.

    Is there any direct evidence that the IMF has varied with time in
our Galaxy?  Globular clusters have been extensively studied to look for
evidence of variability of the IMF, but it is now recognized that the
effects of dynamical evolution are very important for these systems and
that this makes it difficult to derive their initial mass functions. 
In any case, globular clusters might not be representative of all early
star formation, as was noted above, so a better test of variability of
the IMF with time might be to compare the IMF of halo field stars with
that of disk stars.  The Galactic halo has a flatter luminosity function
than the disk (Dahn et al.\ 1995; Reid et al.\ 1996; M\'endez et al.\
1996), but this does not necessarily imply a flatter mass function
because of the different mass-luminosity relations for metal-poor and
metal-rich stars.  When this is taken into account, any evidence that
the halo has a flatter mass function than the disk becomes marginal,
and all that can be said with confidence is that the halo IMF is not
unusually steep.  For example, Richer \& Fahlman (1997) and Chabrier
\& M\'era (1997) find that the slope of the IMF for halo stars is
approximately $x \sim 0.6$ to 0.8, and although this is smaller than
the Salpeter slope $x = 1.35$, it is not significantly smaller than the
slope $x \sim 0.9$ characteristic of the disk IMF in the same mass range
below a solar mass.

    Thus there is at present no clear direct evidence that the IMF in
our Galaxy has varied with time.  The best evidence for a variable IMF
may therefore be provided by the observed chemical abundances in the
stars in our Galaxy and others, and especially by the high heavy-element
abundances in the hot gas in clusters, as will be discussed in the
following section.  Additional evidence may also be provided by recent
observations suggesting high rates of massive star formation in galaxies
at high redshifts, as will be discussed in Section~5.

\section{4~~ EVIDENCE FROM CHEMICAL ABUNDANCES AND SUPERNOVA RATES}

\noindent
    In this section we consider the evidence from chemical abundances
suggesting that the IMF has varied with time in our Galaxy and others,
and we address the question of whether this evidence can be accounted
for with an IMF such as (1) or (2) in which the mass scale \m1\ has
varied with time.  The types of evidence considered here include (1) the
classical G-dwarf problem and the related slow variation of metallicity
with time over most of Galactic history, (2) the chemical abundances
observed in elliptical galaxies and the hot gas in clusters of galaxies,
and (3) the energetics of the hot gas in clusters.

\subsection{4.1~~ Chemical evolution of galaxies and the G-dwarf problem}

\noindent
    The paucity of metal-poor stars in the solar neighborhood relative
to the predictions of simple models of chemical evolution was noted by
van den Bergh (1962) and discussed further by Schmidt (1963), who
suggested that this problem, the classical `G-dwarf problem', can be
solved if the IMF has varied with time in such a way that relatively
more massive stars were formed at earlier times.  A related observation
is the fact that the average metallicity of nearby stars has shown only
a modest increase with time over most of Galactic history, and this
relatively flat `age-metallicity relation' can also be accounted for
with a time-varying IMF; for example, it is reproduced by the
bimodal-IMF model of L86 in which a high-mass mode of star formation
dominates at early times.  However, nearly all recent efforts to model
the chemical evolution of galaxies have assumed instead that the
G-dwarf problem and the flat age-metallicity relation are explained by
the continuing infall of metal-free primordial gas (Larson 1972); most
authors have regarded a variable IMF as an ad hoc hypothesis, while
infall has been considered a more natural solution to the problem
because some amount of continuing infall is predicted by many models of
galaxy formation.  It is therefore relevant to consider again whether
infall is really a satisfactory solution to the G-dwarf problem.

   The original suggestions by Oort (1970) and Larson (1972) that
continuing infall is important for galactic evolution were based on the
assumption that galaxy formation is an inefficient process and leaves
behind a large reservoir of primordial gas that continues be accreted by
galaxies for a long period of time.  However, it now appears that most
of the known intergalactic gas has been heated and enriched in heavy
elements by the effects of early star formation in galaxies; this is
true at least in the large clusters of galaxies that are considered to
be representative samples of the universe.  Thus, there may no longer
be much truly primordial gas, and indeed little evidence has ever been
found for such gas despite many years of searching.  The halo of our
Galaxy contains some `high-velocity clouds' that might represent
infalling (but not primordial) material, yet the total mass in these
clouds is only about $10^7\,$\Msun, and their origin remains unknown
(Wakker \& van Woerden 1997).  Even if they do represent infalling gas,
the estimated infall rate is only a few tenths of a solar mass per year
(Mirabel 1989), an order of magnitude too small to be of major
importance for Galactic chemical evolution.  Infall rates may well have
been much higher at early times, and this could help to solve the
G-dwarf problem, but infall models require large amounts of continuing
infall if they are to account for the flat age-metallicity relation,
especially that indicated by observations of open clusters (Friel 1995).
It is therefore questionable whether infall can by itself adequately
solve the G-dwarf problem and the related flat age-metallicity relation.
It may even be that most galaxies, instead of continuing to accrete
primordial gas, expel significant amounts of supernova-heated and
enriched gas early in their evolution (Section~4.2).

    Another problem for infall models is that the G-dwarf problem
has recently been found to exist also in other galaxies, including
ellipticals (Bressan, Chiosi \& Fagotto 1994; Vazdekis et al.\ 1996;
Worthey, Dorman \& Jones 1996).  Bressan et al.\ (1994) suggest that the
problem can be solved with infall models, but Vazdekis et al.\ (1996)
and Worthey et al.\ (1996) consider infall models to be unlikely for
elliptical galaxies and suggest instead that a variable IMF may account
for the observations.  There is less time for infall to be effective in
elliptical galaxies than in spirals, and in any case there is evidence
that the elliptical galaxies in clusters lose rather than accrete large
amounts of gas (Section~4.2).  One might also expect the importance of
infall effects to depend on the environment, yet the G-dwarf problem
seems to exist in all galaxies studied, independently of their
environments; this suggests that the origin of the G-dwarf problem may
be an intrinsic feature of galactic evolution such as a time-varying
IMF.

    A time-varying IMF can solve the G-dwarf problem if relatively few
low-mass stars were formed during the early period when the metallicity
was low.  Since simple models of chemical evolution are based on the
instantaneous recycling approximation, which applies best to elements
like oxygen that are produced only in massive stars, it is best to use
oxygen abundances when comparing predicted and observed abundance
distributions.  Studies of the oxygen abundance distribution of nearby
stars have consistently shown that only about 5--10\% of the stars in
the solar neighborhood have oxygen abundances below half of the median
value (Pagel 1989; Sommer-Larsen 1991; Wyse \& Gilmore 1995; Rocha-Pinto
\& Maciel 1996), representing a deficiency by a factor of 3 or more
relative to the simple-model prediction that 28\% of the stars should
have abundances in this range.  These studies give the abundance
distribution of all stars in a vertical column through the local
Galactic disk, so they show that even at large distances from the
Galactic plane there are not enough oxygen-poor stars to be consistent
with a simple model.  The exact magnitude of the deficiency depends
somewhat on the assumed relation between oxygen abundance and iron
abundance; for example, if the usual assumption that ${\rm [O/H]} =
0.5{\rm [Fe/H]}$ is adopted, the fraction of stars in a local column
having oxygen abundances below half of the median value is about 6\%
according to Wyse \& Gilmore (1995) and 4\% according to Rocha-Pinto
\& Maciel (1996), while if it is assumed instead that ${\rm [O/H]} = 0.65{\rm [Fe/H]}$, as may be more consistent with recent evidence
(McWilliam 1997), these numbers become 12\% and 9\%, respectively.

    If we assume, accordingly, that only 10\% or less of the stars in a
local column have oxygen abundances below half of the median value, this
can be explained if the number of low-mass stars that formed when the
oxygen abundance was this low was reduced by a factor of 3 relative to
the number of massive oxygen-producing stars formed.  With an IMF of
form (1), such a reduction can be achieved if the peak mass \mp\ was
increased by a factor of 4 to 1$\,$\Msun\ during the early oxygen-poor
period; the number of 0.9$\,$\Msun\ stars that formed during this
period is then reduced by a factor of 2.9 relative to the number of
20$\,$\Msun\ stars formed, the reduction factor being larger for stars
less massive than 0.9$\,$\Msun.  With an IMF of form (2), the same
reduction in the ratio of 0.9$\,$\Msun\ stars to 20$\,$\Msun\ stars is
obtained if the mass scale \m1\ was increased by a factor of 6 at early
times.  Since in this simple model only 10\% or less of the presently
observed low-mass stars were formed with a top-heavy IMF, the effect of
this change in the IMF on the present luminosity function of nearby
stars will be small.

    The above discussion assumes that the amount of oxygen produced by
massive stars is independent of their metallicity, but this may not be
true because massive stars lose much of their mass in winds driven by
radiation pressure, and this can cause the amount of oxygen produced
to decrease with increasing metallicity (Maeder 1992).  A decrease in
oxygen yield with increasing metallicity might help to solve the G-dwarf
problem, but the effect estimated by Maeder is less than a factor of~2,
and there is in any case a compensating effect with the opposite sign,
namely that more oxygen can end up in black holes in stars of lower
metallicity (Woosley \& Weaver 1995).  As presently estimated, these
effects do not appear to be large enough to have an important impact on
the G-dwarf problem, but they probably make any quantitative prediction
of oxygen yields uncertain by at least a factor of~2.

    Are models with a time-varying IMF consistent with other
constraints?  A general prediction of variable-IMF models is that they
predict more mass in stellar remnants, mostly white dwarfs, than do
standard models.  If the mass of a stellar remnant is the larger of
0.6$\,$\Msun\ and 0.15 times the initial stellar mass, as seems
consistent with present evidence, the model of Schmidt (1963) predicts
that the surface density of mass in remnants in the local Galactic disk
is about 50$\,$\Msun$\,$pc$^{-2}$, while the model of L86 predicts a
surface density of about 30$\,$\Msun$\,$pc$^{-2}$ in remnants and the
above simple two-stage model predicts about 12$\,$\Msun$\,$pc$^{-2}$ in
remnants.  The model of Schmidt (1963) and probably the model of L86 are
therefore now excluded by the fact that the dynamically determined total
surface density of matter within a kiloparsec of the Galactic midplane,
including halo dark matter, is only about 70$\,$\Msun$\,$pc$^{-2}$,
about 40$\,$\Msun$\,$pc$^{-2}$ of which is already accounted for by
visible stars and gas (Kuijken \& Gilmore 1991).  However the above
simple model assuming only the minimum change in the IMF needed to solve
the G-dwarf problem is not excluded by this constraint, and in fact it
does not differ much from a standard model in this respect, predicting
only about 20\% more mass in remnants.  Thus a variable-IMF model that
solves the G-dwarf problem does not necessarily predict much more mass
in remnants than a standard model, and is not excluded on this basis.

    Another constraint may be provided by the oxygen production rate
needed to achieve agreement with observed oxygen abundances.  In the
above simple model, the total amount of oxygen produced for each
presently visible 0.9$\,$\Msun\ star is about 23\% larger than in a
standard model.  However, this difference is much smaller than the
uncertainty in the predicted oxygen yields for massive stars (Gibson
1997), so it does not significantly constrain such a model.  The more
extreme bimodal-IMF model of L86 overproduces oxygen by about a factor
of 3 compared with a standard model, and this may exclude that model. 
However, it should be noted that stellar abundances do not constrain the
IMF at all if galaxies lose large amounts of enriched gas to the
intergalactic medium, as has apparently happened in clusters of galaxies
(see below).  The only firm constraint on an early top-heavy IMF is then
that provided by the mass left behind in stellar remnants.

    We conclude that a time-varying IMF can solve the G-dwarf problem
without requiring large departures from a standard model in other
respects, and without violating any of the constraints that have
been discussed.  The main possible problem with models assuming a
time-varying IMF is that there is no clear evidence that the IMF of the
oldest stars in our Galaxy, those in the halo, differs significantly
from that of the disk; therefore, if the G-dwarf problem is to be solved
with a time-varying IMF, either early star formation must have produced
a more top-heavy IMF in the disk than in the halo, or the Galactic disk
must have been pre-enriched by a population of massive stars unrelated
to the visible halo population.  An early epoch of preferentially
massive star formation could have coincided with the period when the
Galactic thick disk was formed, since the thick disk contains about
10\% of the stars in a local column; at present there are no constraints
on the IMF of the thick disk.  The possibility that metal-free
`population~III' stars could have pre-enriched the Galactic disk (Truran
\& Cameron 1971) will be discussed in Section~7.  Meanwhile, we note
that it remains true that none of the explanations that has been
proposed for the G-dwarf problem, including gas flows and inhomogeneous
chemical evolution, can yet be definitely ruled out, and that all of
them could still play some role in its solution (Tinsley 1980).

\subsection{4.2~~Abundances in elliptical galaxies and the hot gas in
            clusters}
\noindent
    Since elliptical galaxies are generally older in their stellar
content than spirals, they might be expected to show more evidence for
a variable IMF\null.  The observed properties of elliptical galaxies do,
in fact, show several trends that might reflect a varying IMF\null.  One
such trend is the `fundamental plane' correlation among the effective
radius, surface brightness, and velocity dispersion of elliptical
galaxies, which implies a systematic increase of mass-to-light ratio
with mass (Guzm\'an, Lucey \& Bower 1993; J\o rgensen, Franx \&
Kjaergaard 1996).  Similar trends are found quite generally in
early-type galaxies (Burstein et al.\ 1997).  In addition, both the
magnesium abundance Mg/H and the ratio of magnesium to iron abundances
Mg/Fe increase systematically with mass among early-type galaxies (Faber
1977; Worthey, Faber \& Gonz\'alez 1992).  All of these trends could in
principle be caused by a varying IMF, but the increase in Mg/Fe with
mass could also reflect a shorter formation timescale for the more
massive galaxies (Worthey et al.\ 1992), as might perhaps be expected
since current cosmological simulations suggest that the biggest galaxies
tend to form first.  The similar variations in the ratio of
alpha-elements to iron that are seen in other old stellar populations
are all well explained by the expected larger contribution of type~II
supernova products in older populations (McWilliam 1997), so a similar
explanation may apply also to the Mg/Fe ratio in ellipticals.  The
remaining trends that might still reflect a varying IMF are then the
systematic increases of both mass-to-light ratio and metallicity with
mass among early-type galaxies.  Other explanations of these trends are
also possible (e.g., Guzm\'an et al.\ 1993; see below), but we consider
first whether a variable IMF of the type discussed above can account for
them.

    It is noteworthy that the mass-to-light ratio \M/L\ and the
magnesium abundance Mg/H both increase with mass among early-type
galaxies, since this is the opposite of what would be predicted if the
IMF were a simple power law with a variable slope; the mass would then
mostly be in low-mass stars and \M/L\ would anti-correlate with Mg/H. 
However, the observed correlation can be accounted for if the IMF is
bimodal and if stellar remnants produced by an early high-mass mode of
star formation contribute importantly to the  mass; then, if the
high-mass mode is more dominant in the more massive galaxies, both \M/L\
and Mg/H will increase with mass (L86).  A model of this type has been
developed by Zepf \& Silk (1996, hereafter ZS), who show that it can
account not only for the increase of \M/L\ with mass among elliptical
galaxies but also for the large amount of mass in heavy elements that
is observed in the hot gas in clusters of galaxies.  In the ZS model
the heavy elements in the hot gas are assumed to be produced by the
high-mass mode of star formation, suggested by these authors to be
associated with merger-induced starbursts in forming elliptical
galaxies.

     The predictions of the ZS model do not depend strongly on the
assumption that the IMF is bimodal, but only require the IMF to be more
top-heavy in the more massive galaxies.  This would be true also for a
single-peaked IMF such as (1) if the peak mass \mp\ is higher in the
more massive galaxies, as might occur if the more massive galaxies form
earlier than the less massive ones.  The total observed increase of
\M/L\ with mass is about a factor of 4, and an increase of this
magnitude would be predicted if the peak mass in IMF (1) were increased
from 1$\,$\Msun\ to 2.5$\,$\Msun.  Values of \mp\ in this range do not
seem implausible for star formation at large redshifts, as was noted in
Section~3.  However, the increase in \M/L\ that can be obtained in this
way is very sensitive to the assumed form of the lower IMF; with an IMF
of form (2), for example, \M/L\ increases by a much smaller factor of
only 1.2 for the corresponding increase in the mass scale \m1.  Thus
an IMF that falls off steeply at the low end is required to obtain
a sufficiently large variation in \M/L.  The model of ZS satisfies
this requirement by assuming that the IMF of the high-mass mode is
sharply truncated below 3$\,$\Msun.

    For a given mass in remnants, a model with a single-peaked IMF
predicts a similar mass of heavy elements as does the ZS model.  For
example, if the total mass in remnants is $9 \times 10^{12}\,{\rm
M}_\odot$, as in the ZS model of the Coma cluster (for $h = 0.65$),
that model predicts that the mass of iron produced is about $1.2 \times
10^{11}\,{\rm M}_\odot$, in satisfactory agreement with the observed
iron content of $7 \times 10^{10}\,{\rm M}_\odot$; for IMF (1) with
$m_{\rm p} \sim 2\,{\rm M}_\odot$, the amount of iron produced with the
same stellar production rate (Woosley \& Weaver 1995) is about $8 \times
10^{10}\,{\rm M}_\odot$, again in satisfactory agreement with the
observations.  This agreement is not very sensitive to the assumed value
of \mp, and it shows that a smoothly peaked IMF of form (1) can account
for the observed amount of heavy elements in the hot gas of the Coma
cluster as well as can the bimodal IMF of ZS.

    Many efforts to model in more detail the properties of elliptical
galaxies and the hot gas in clusters have also concluded that the IMF
in the proto-ellipticals must have been top-heavy (Elbaz, Arnaud, \&
Vangioni-Flam 1995; Loewenstein \& Mushotzky 1996; Gibson 1997; Gibson
\& Matteucci 1997b), and possibly more top-heavy in the more massive
galaxies (Matteucci 1994; Angeletti \& Giannone 1997).  Most of these
studies have tried to model the chemical evolution of elliptical
galaxies with specific galactic wind models, and most have assumed that
the IMF is a power law with an adjustable slope, in which case a slope
in the range $x \sim 0.8$ to 1.0 is favored.  However, the same
observations can be fitted equally well with an IMF of form (1) or (2)
in which the mass scale \m1\ is larger than in the standard present-day
IMF\null.  The strongest constraint on the early IMF in clusters of
galaxies, which does not depend on the details of galactic evolution
models, is provided by the large total mass of heavy elements observed,
mostly in the hot gas.  Typically, the total mass of heavy elements is
a few times that predicted by standard models based on a Salpeter IMF;
in the Coma cluster, for example, the ratio of heavy-element mass to
total cluster luminosity is about 3 times what would be predicted
for a Salpeter IMF\null.  An increase by a factor of 3 in the amount
of heavy elements produced relative to the number of low-mass stars
formed is the same effect that was needed above to solve the G-dwarf
problem, and it can be achieved with the same kind of change in the IMF
at early times, for example by an increase in the mass scale \m1\ by
a factor of 4 for IMF (1).  This has approximately the same effect as
changing the slope $x$ of a power-law IMF by $-0.4$, but an IMF of form
(1) or (2) with a variable mass scale may be more consistent with the
evidence mentioned in Section~2 suggesting that the IMF has a universal
Salpeter-like form for large masses while it flattens or declines at
the low end.

     A finer point concerns the relative abundances of the various heavy
elements that can now be measured in the hot gas (Mushotzky et al.\
1996; Loewenstein \& Mushotzky 1996).  These and previous studies have
generally found that alpha-elements such as oxygen and silicon are
overabundant relative to iron, as would be expected if the hot gas
had been enriched mainly by type~II supernovae.  However, when the
$\alpha$/Fe ratios are corrected for a recent downward revision of the
solar iron abundance, they become only slightly higher than solar and
suggest a significant contribution of iron from type Ia supernovae
(Ishimaru \& Arimoto 1997).  This conclusion has been questioned by
Gibson, Loewenstein \& Mushotzky (1997) because of the uncertainties
that remain in the predicted heavy-element yields, but $\alpha$/Fe
ratios that are only slightly above solar and that reflect a significant
contribution from type~Ia supernovae would in fact be expected if the
IMF always has a Salpeter-like slope at the upper end like IMFs (1) or
(2), since the predicted ratio of type~Ia to type~II supernovae is then
nearly the same as for a standard IMF\null.  Note that the abundance
ratios observed in the hot gas differ, as would be expected, from those
in the stars, since the stars are enriched only by early supernovae that
are mostly of type~II, so that the stars show enhanced $\alpha$/Fe
ratios, while the hot gas is enriched also by later type~Ia supernovae
and thus shows nearly solar abundance ratios (Renzini 1997). 
Loewenstein \& Mushotzky (1996) note that there is marginal evidence
from some of the element ratios favoring a Salpeter slope over a smaller
slope for the massive stars, and thus favoring an IMF of form (1) or (2)
over a simple power law with $x \sim 0.9$ at all masses.  Wyse (1997)
questions the need for any nonstandard IMF and suggests that most of
the cluster data are consistent with a universal Salpeter-like IMF,
but notes nevertheless that the high overall abundances of heavy
elements observed in some clusters may be difficult to reconcile with
a universal IMF.

     In summary, we conclude that the chemical properties of the
galaxies and the hot gas in clusters can be accounted for satisfactorily
if the early IMF has a form similar to (1) or (2) with a mass scale
\m1\ of the order of 1--2$\,$\Msun.  An IMF of this form may be more
consistent with the observed abundance ratios than a simple power-law
IMF with $x \sim 0.9$ at all masses, which otherwise could account
equally well for the overall enrichment in heavy elements.  The increase
of \M/L\ with mass among elliptical galaxies might also be explainable
if the mass scale \m1\ increases with galactic mass, but only if the IMF
falls off steeply at the low end like IMF (1), i.e.\ more steeply than
is suggested by most of the evidence concerning the IMF in our Galaxy. 
Thus the increase in \M/L\ with mass among elliptical galaxies is not as
easily accounted for with a varying IMF as are the chemical abundances
in galaxies and clusters, since more extreme assumptions are required. 
The alternative possibility that the increase of \M/L\ with mass results
from a decrease in the spatial concentration of the visible matter
relative to the dark matter in elliptical galaxies may therefore be
a more promising explanation of the observed trend (Guzm\'an et al.\
1993); as was noted by these authors, a decrease in the relative spatial
concentration of the visible matter and hence an increase in the
measured mass-to-light ratio with increasing galaxy mass would be
expected if mergers play an increasingly important role in the formation
of the more massive galaxies.

     The chemical abundances in the hot gas can of course only constrain
the IMF averaged over an entire cluster and not the IMF in any
individual type of galaxy.  The question of whether dwarf galaxies
contribute importantly to the enrichment of the hot gas (Trentham 1994,
Gibson \& Matteucci 1997a) is therefore not relevant to this integral
constraint.  The essential property of cluster galaxies causing them to
have a systematically top-heavy IMF may simply be that as a group, they
are dominated by very old stellar populations that probably contain the
first visible stars to be formed in the universe.  If the mass scale
for star formation was indeed generally higher at early times, then
clusters of galaxies could plausibly have had an average IMF that was
systematically more top-heavy than that found, for example, in the local
disk of our Galaxy or in the Magellanic Clouds.

\subsection{4.3~~Energetics of the hot gas in clusters}

\noindent
    If the early star formation in clusters of galaxies had a top-heavy
IMF, this would imply not only more heavy-element production but also
more energy input to the hot gas from supernovae than is predicted by
standard models.  Since the total mass of heavy elements per unit
luminosity in clusters is typically a few times that predicted for a
Salpeter IMF, the number of supernovae that occurred at early times must
also have been a few times larger than in standard models.  As discussed
by Elbaz et al.\ (1995) and Loewenstein \& Mushotzky (1996), this means
that a larger amount of energy would have been available to drive
galactic winds, and suggests that a substantial fraction, perhaps half,
of the initial mass of the cluster galaxies could have been ejected in
such winds.  The fact that most of the heavy elements in clusters are
in the intracluster medium rather than in the galaxies implies that the
cluster galaxies indeed lost most of their supernova-enriched gas, in
contrast to conventional models where only a modest fraction of the gas
is lost (Larson \& Dinerstein 1975).

    These energetic outflows can contribute to heating the intergalactic
medium, and such extra heating may be required because the hot gas in
clusters has a more extended spatial distribution than the galaxies or
the dark matter and more energy per unit mass than the galaxies (David,
Jones \& Forman 1995, 1996).  According to these authors, the thermal
energy per unit mass of the hot gas is typically about twice the
kinetic energy per unit mass of the elliptical galaxies, and this
implies an excess thermal energy of the order of $4 \times 10^{49}$ ergs
per \Msun\ of visible stars.  If this energy comes from supernovae, and
if the energy supplied per supernova is equal to the explosion energy of
$10^{51}$ ergs, then one supernova is required for every 25$\,$\Msun\
of visible stars, about 3 times the number predicted for a standard
IMF\null.  If significant supernova energy is lost by radiative cooling,
as is expected (Larson 1974), then even more supernovae are needed. 
Thus the number of supernovae needed to heat the gas may exceed the
number predicted for a standard IMF by a significant factor, again
suggesting a top-heavy early IMF. 

    If supernovae can indeed supply a thermal energy comparable to the
binding energy of the gas in a cluster of galaxies, then large amounts
of gas may be lost not only from individual galaxies but also from
entire groups and small clusters, and only the largest clusters (if any)
may have been able to retain all of their enriched gas.  Both the gas
content and the heavy-element content of groups and clusters increase
with cluster size (David et al.\ 1995), and this suggests that most
groups and small clusters have indeed lost most of their gas and that
only the largest ones have evolved as closed systems (David 1997;
Renzini 1997).  If this is true, then a significant fraction of the
heavy elements ever produced in the universe may have been ejected from
galaxies into a hot intergalactic medium that is not bound in clusters
and has not yet been observed.  Although present estimates of the
history of massive star formation in the universe seem consistent with
its known heavy-element content, a significant amount of early star
formation could have been missed in these estimates because of the
neglect of dust extinction (see Section~5); in this case a top-heavy
early IMF may be required quite generally, even outside clusters, and
galaxies might have produced more heavy elements than are predicted by
standard models and might have ejected a significant fraction of their
heavy elements into an intergalactic medium.

    Thus the energetics of the hot gas in clusters suggests, in
agreement with the heavy-element content, that the early supernova rate
in clusters may have been at least 3 times that predicted by standard
models, implying that the early IMF produced at least 3 times more
massive stars per solar-mass star than a Salpeter IMF\null.  The
detailed shape of the early IMF is not constrained by this evidence,
which could be accounted for equally well by a power-law IMF with
$x \sim 0.9$, an IMF of form (1) or (2) with $m_1 \sim 1.5\,$\Msun,
or a bimodal IMF\null.  However, an IMF similar to (1) or (2) seems
preferable for the reasons mentioned above, including the fact that it
is more consistent with the form of the present-day IMF and the fact
that it yields better consistency with the observed abundance ratios
in clusters than a simple power law.

\section{5~~LUMINOSITY EVOLUTION OF GALAXIES}

\noindent
    If the old stars responsible for most of the present luminosity of
early-type galaxies were formed with a top-heavy IMF, these galaxies
will fade more rapidly with time than would be the case with a standard
IMF\null.  According to Tinsley (1980), the luminosity of an old stellar
system varies with time as $L \propto t^{-1.3+0.3x}$; thus if $x$
changes by $-0.5$, the exponent of $t$ changes by only $-0.15$ and the
effect may be difficult to measure, while if the value of $x$ becomes
as small as 0 at $m \sim {\rm M}_\odot$, as in some of the examples
considered above, the exponent of $t$ changes by $-0.4$ and the effect
may become measurable, implying an extra dimming by a factor of 2 in
luminosity over a factor of 5 in cosmic time.  For actively star-forming
galaxies whose light is dominated by massive stars, the effect of a
top-heavy IMF on luminosity evolution can be much more dramatic,
especially if the star formation rate declines strongly with time;
if the IMF contains relatively few low-mass stars, a system that is
initially extremely luminous can then evolve into one that is extremely
faint.

    Direct studies of galaxy evolution by observations at a range of
redshifts are still at an early stage, but the first results of such
studies suggest that a relatively flat early IMF may be required to
account for the data.  Lilly et al.\ (1996) have used data from the CFRS
redshift survey to determine the evolution of the comoving luminosity
density of the universe since a redshift of $z = 1$, and they find
strong evolution that is compatible with simple models of galaxy
evolution based on a Salpeter IMF but incompatible with models based
on the steeper Scalo (1986) IMF\null.  If dust extinction is important,
these results may require an IMF that is even flatter than the Salpeter
IMF\null.  Totani, Yoshii \& Sato (1997) find that the Lilly et al.\
(1996) results are inconsistent with the galaxy evolution models of
Arimoto \& Yoshii (1987), again based on a Salpeter-like IMF, unless a
large cosmological constant is assumed, but an alternative explanation
for the discrepancy would be a flatter IMF\null.  Rapid fading due to a
top-heavy IMF might also help to account for the apparent disappearance
of many `faint blue galaxies' that are abundant at intermediate
redshifts but that appear to have no equally abundant counterparts at
low redshift (Ellis 1997; Ferguson \& Babul 1998).

    Madau et al.\ (1996), Guzm\'an et al.\ (1997), and Madau, Pozetti \&
Dickinson (1998) have used observations of high-redshift galaxies in the
Hubble Deep Field to extend the results of Lilly et al.\ (1996) to a
redshift $z \sim 4$, and they find again that the evolution of the
luminosity density of the universe at various wavelengths is compatible
with models of galaxy evolution based on a Salpeter IMF but incompatible
with models based on the steeper Scalo IMF, and possibly incompatible
even with a Salpeter IMF if extinction is sufficiently important.  The
neglect of dust extinction is questionable since actively star-forming
galaxies at low redshifts generally show heavy obscuration and radiate
much of their energy at infrared wavelengths, and this may be true also
for star-forming galaxies at larger redshifts (Mazzei \& De Zotti 1996;
Rowan-Robinson et al.\ 1997).  Smail, Ivison \& Blain (1997) have
discovered an abundant class of lensed submillimeter sources at
redshifts $z > 1$ that could be heavily reddened young galaxies, and
they suggest that a large upward revision may be needed in the cosmic
rate of massive star formation at high redshifts estimated by Madau et
al.\ (1996).  In discussing the interpretation of these results, Blain
et al.\ (1998) conclude that unless obscured AGNs account for a major
fraction of the observed flux, a top-heavy early IMF is probably
required in order not to produce too many low-mass stars that would
remain visible to the present time.

    Recently, a far-infrared background radiation has been detected
in the COBE data that could come from heavily obscured, actively
star-forming galaxies at high redshifts (Puget et al.\ 1996; Guiderdoni
et al.\ 1997; Schlegel, Finkbeiner \& Davis 1998; Hauser et al.\ 1998). 
The background radiation intensity reported by Hauser et al.\ (1998)
is similar to the level predicted by ZS from their model invoking an
early high-mass mode of star formation to account for the abundance of
heavy elements in clusters, and therefore this observation is consistent
with, and may provide support for, models with a top-heavy early
IMF\null.  In discussing possible interpretations of the far-IR
background, Dwek et al.\ (1998) show that all of the models that have
been proposed to describe the cosmic history of star formation that are
consistent with low-redshift observations fail by at least a factor of 2
to account for the observed background intensity, but they show that it
can be accounted for with a model that includes a period of formation
of exclusively massive stars at high redshifts, as was was proposed in
the models of Elbaz et al.\ (1995) and ZS that postulate a bimodal IMF.
Again, we note that the detailed form of the IMF is not constrained by
this evidence, which could be accounted for equally well with an IMF
similar to (1) or (2) in which the mass scale \m1\ was higher at early
times.

    Thus, while studies of high-redshift galaxies and infrared
background radiation do not yet provide conclusive evidence for
time-variability of the IMF, they have the potential to test in a
relatively direct way whether the global IMF has varied with time, and
some recent data suggest that such variability has indeed occurred. 
Further observations capable constraining the cosmic history of massive
star formation will therefore be of great interest for constraining the
time evolution of the IMF.

\section{6~~DARK MATTER AND MACHOS}

\noindent
    A question of current interest is whether the remnants of large
numbers of massive stars formed at early times could contribute
significantly to the dark matter in galactic halos (Larson 1987; Burkert
\& Silk 1997).  Evidence that this might be the case has recently been
provided by the claimed detection via gravitational microlensing of
stellar-mass `machos' in the halo of our Galaxy (Alcock et al.\ 1997);
these authors estimate that half of the dark mass in the Galactic halo
out to a radius of 50$\,$kpc could be in the form of machos with a
typical mass of $\sim 0.5\,$\Msun.  The only known dim objects of this
mass are white dwarfs, but if the estimated typical macho mass increases
as microlensing events of longer duration are recorded, neutron stars
and black holes might also become candidates.  All of these types of
objects could in principle exist as the remnants of large numbers of
massive stars formed at early times in the Galactic halo.  Several
authors have considered the viability of such explanations of the macho
results and have concluded that they are strongly constrained by the
scarcity of visible low-mass halo stars and by the lack of evidence for
the expected heavy-element enrichment of the Galaxy, implying an IMF
that is sharply peaked at a few \Msun\ and falls off very steeply at
both lower and higher masses; moreover, any significant white-dwarf
component must be very old in order not to conflict with observed white
dwarf numbers (Adams \& Laughlin 1996; Flynn, Gould \& Bahcall 1996;
Kawaler 1996; Chabrier, Segretain \& M\'era 1996; Graff, Laughlin \&
Freese 1998).  The constraint on the upper IMF is relaxed if most of
the heavy elements produced by the massive stars are ejected into an
intergalactic medium, possibly by starburst activity associated with
the formation of the halo, as suggested by Fields, Mathews \& Schramm
(1997); these authors propose a model for halo formation with a
lognormal IMF that peaks at 2.3$\,$\Msun, and they claim that it can
account for $\sim\,$50\% of the dark mass as remnants. This model is
qualitatively similar to the models with a peaked IMF considered in
Section~4.2 to describe the early evolution of elliptical galaxies in
clusters, where there is indeed evidence that most of the heavy elements
are ejected into an intergalactic medium; therefore it is of interest to
consider whether similar models can account for significant amounts of
dark matter as remnants.

    If we consider an IMF of form (1) as a possible description of early
star formation in the Galactic halo, an old population with an IMF of
this form has a mass-to-light ratio of about 17 if the peak mass \mp\ is
2.5$\,$\Msun, 120 if \mp\ is 4$\,$\Msun, and 1800 if \mp\ is 6$\,$\Msun;
the associated mass is mostly in white dwarfs for the smaller value of
\mp\ and mostly in neutron stars and black holes for the larger values. 
Thus, it is possible in principle to account for large amounts of dark
matter with an IMF of this form if the peak mass \mp\ is high enough. 
The predicted amount of dark mass could be even larger if metal-poor
stars lose less mass in winds than metal-rich stars (Maeder 1992) and
eject less mass in supernova explosions (Woosley \& Weaver 1995),
leaving more mass behind in remnants.  However, the possibility of
achieving a very large \M/L\ in this way depends critically on having an
IMF that falls off sufficiently steeply at low masses, as has already
been noted (Section~4.2); with an IMF of form (2), for example, the
predicted values of \M/L\ for the cases considered above are only about
6, 7, and~8 respectively.  A sharp truncation of the lower IMF could of
course yield an arbitrarily large \M/L, but this does not seem plausible
physically, and even the smooth exponential falloff of IMF (1) at low
masses is steeper than is suggested by the available evidence concerning
the local IMF\null.  The possibility that population~III stars might
have had a sufficiently strongly truncated lower IMF to allow the
existence of large amounts of dark matter in the remnants of such stars
will be discussed in Section~7.

     If the galaxies in clusters have as much associated dark matter in
the form of machos as has been suggested to exist in the halo of our
Galaxy (Alcock et al.\ 1997), such machos would make a significant
contribution to the cluster mass.  David (1997) has noted that if half
of the dark matter associated with cluster galaxies out to a radius
of 50$\,$kpc is in the form of stellar remnants, then the baryonic
content of clusters could be as large as $\sim\,$50\%, with remnants
contributing $\sim\,$10--20\% of the total.  Most of the hot gas in
clusters might then have been processed through the massive remnant
progenitors.  However, the total amount of mass in remnants would then
be 10--20 times larger than in the ZS model discussed in Section~4.2,
in which remnants make up only about 1\% of the cluster mass; the amount
of heavy elements produced would therefore also be larger by a factor
of 10--20 than in the ZS model, and larger by a similar factor than the
amount observed, assuming the same production rates in massive stars. 
Thus, such a model is not viable unless nearly all of the heavy elements
produced by the massive stars are either lost from the cluster, which
seems unlikely, or trapped in black-hole remnants.  Very massive
population~III stars are predicted to collapse completely into black
holes (Section~7), but such black holes would be much more massive than
the machos detected so far in the Galactic halo, and therefore could not
account for the macho results.

     We conclude, as have previous authors, that it is difficult with
any plausible assumptions to construct models that can account for a
large fraction of the dark matter in the universe as being in low-mass
stellar remnants such as white dwarfs, neutron stars, or small black
holes.  Some small fraction of the dark matter, perhaps a few percent,
could plausibly be in this form, but the only apparent way to account
for large amounts of dark matter as stellar remnants would be with the
massive remnants of very massive population~III stars.

\section{7~~POPULATION III STARS}

\noindent
     When the first stars formed, no heavy elements had yet been
produced and recycled into new stars by supernovae, so these first
`population~III' stars (Carr 1994) would have contained no heavy
elements.  The complete absence of heavy elements leads to qualitative
differences in star formation and stellar evolution that can have
important consequences for the stellar IMF and heavy-element production.
Woosley \& Weaver (1995) have noted, for example, that metal-free stars
develop much more compact cores than stars with finite metallicity,
with the result that less of the core mass is ejected in a supernova
explosion and more ends up in a black hole; this reduces the
heavy-element production rate by about a factor of 4 for a standard
IMF\null.  Thus population~III stars might be better candidates for
dark-matter progenitors than stars with finite metallicity, both because
of the predicted higher remnant masses and because of the reduced
heavy-element production.  However, an even more important difference
may be that in the complete absence of heavy elements, star formation
almost certainly produces a very different and much more top-heavy IMF
(see below).

    Would one expect significant numbers of population~III stars to have
formed?  The failure of astronomers to find any metal-free stars despite
many years of searching has long been a puzzle, and the explanation
usually offered is that very few such stars were needed to produce the
very small amount of heavy elements observed in the most metal-poor
stars.  However, it is not very plausible that the number of metal-free
stars formed should have been negligibly small.  The first stars with
finite metallicity could only have appeared after the first metal-free
stars had completed their evolution and exploded as supernovae, and
after the resulting hot supernova ejecta had cooled, mixed with the
surrounding gas, and become incorporated into new collapsing clouds and
stars.  The lifetimes of massive stars are at least a few million years,
and the time required for their supernova ejecta to be recycled into new
stars is probably at least of the same order; at present, this recycling
time is of the order of tens of millions of years (Larson 1996).  Thus,
the time interval between the formation of the first metal-free stars
and the formation of the first finite-metallicity stars was probably not
a negligible fraction of the age of the universe when star formation
began; at a minimum, it must have been at least 1\% of the age of the
universe at that time.  If stars continued to form during this interval,
an appreciable number of metal-free stars could have formed.

    If population~III stars did form in significant numbers, they must
have had an IMF that contained essentially no solar-mass stars, since no
metal-free solar-mass stars have yet been seen.  An IMF strongly biased
toward high masses would probably be expected for population~III stars,
since in the absence of heavy elements the only available coolant in
primordial clouds is molecular hydrogen, which cannot cool the gas much
below about 100$\,$K.  A number of recent studies of the thermal
properties of primordial gas clouds have predicted temperatures in the
range $\sim\,$100--1000$\,$K that are one to two orders of magnitude
higher than the temperature in present-day star-forming clouds (Anninos
\& Norman 1996; Haiman, Thoul \& Loeb 1996; Tegmark et al.\ 1997).  The
Jeans mass at a given pressure would therefore have been 2 to 4 orders
of magnitude higher in such primordial clouds than in present-day
star-forming clouds.  If the pressures in the primordial clouds were
not very much higher than those in present-day clouds, the Jeans mass
for primordial star formation would then have been several orders of
magnitude higher than that characterizing present-day star formation.

    A further important result is the demonstration by Uehara et al.\
(1996) that if molecular hydrogen is the only coolant, the minimum
fragment mass that can be attained before high opacity to the cooling
radiation prevents further fragmentation is approximately the
Chandrasekhar mass.  From a more detailed calculation, Nakamura \&
Umemura (1998) find that this minimum mass is about 4$\,$\Msun, and they
suggest that most population~III stars formed with masses larger than
16$\,$\Msun.  The minimum fragment mass set by opacity is much smaller
than the Jeans mass, and it is a strong lower limit that is not
sensitive to details of the star formation process such as the geometry
of the fragmenting cloud.  Under present conditions, this lower limit is
about 0.007$\,$\Msun\ (Low \& Lynden-Bell 1976; Rees 1976; Silk 1977),
but few if any stellar objects with such a small mass have been found,
and the great majority of stars have much larger masses.  Thus, the
opacity limit may in practice rarely be reached, and most stars may
form with much larger masses that are closer to the Jeans mass.  These
considerations suggest that both the characteristic mass of the IMF and
the minimum stellar mass were shifted upwards by several orders of
magnitude for population~III stars; most of these stars might then have
had masses of hundreds or thousands of solar masses, while none had
masses as small as 1$\,$\Msun. Thus, significant numbers of massive
population~III stars could have existed without conflicting with the
observed absence of solar-mass metal-free stars.

    Can the remnants of population~III stars contribute importantly to
the dark matter in the universe?  For this to be possible, several
strong requirements must be satisfied:  (1) Early star formation must
have been very efficient, and must have processed most of the baryonic
content of the universe into massive stars.  Present-day star formation
is very inefficient, but the first star formation might have been more
efficient if the negative feedback effects that now limit the rate of
star formation were less important; radiation pressure and stellar
winds, for example, might have been much less important in the absence
of heavy elements.  (2) If most of the stellar mass is to be converted
into dark matter, these massive stars must then collapse with high
efficiency into dark objects, presumably mostly black holes.  Moreover,
if heavy elements are not to be greatly overproduced, most of the heavy
elements synthesized in these stars must also end up in black holes. 
These requirements might be satisfied for `very massive objects' with
masses larger than 200$\,$\Msun, which are predicted to collapse
completely into black holes (Carr 1994).  Another possibility might be
that some massive stars undergo secular core contraction to a black hole
and thus `fizzle' without producing a supernova event (Hayashi, Eriguchi
\& Hashimoto 1998).  (3) If limits on the cosmic background radiation at
various wavelengths are not to be exceeded, the formation of large
numbers of massive population~III stars is constrained to occur at very
large redshifts of the order of 30 or more (Carr 1994).  Currently
popular variants of standard CDM cosmology do not predict large amounts
of star formation at such high redshifts, but significant star formation
might occur at such redshifts in some less conventional cosmological
models that have been proposed, such as a purely baryonic universe
(Peebles 1987, 1993; Gnedin \& Ostriker 1992; Cen, Ostriker \& Peebles
1993).

    Given the number of requirements that must be satisfied, it does not
presently seem very likely that the remnants of massive population~III
stars can be major contributors to the dark matter.  However, even if
they do not contribute importantly to explaining machos or dark matter,
population~III stars might still play an important role in pre-enriching
the gas in galaxies and helping to solve the ubiquitous G-dwarf problem.
If such stars produce enough oxygen to pre-enrich the gas in the disk of
our Galaxy to half of the present median stellar oxygen abundance, then
the total mass left behind in their remnants, assuming a standard upper
IMF and the oxygen production rates of Woosley \& Weaver (1995) for
metal-free stars, will be about $5 \times 10^9\,$\Msun, which is a small
fraction of the mass of the Galactic disk and an even smaller fraction
of the halo mass.  Thus the predicted amount of mass in remnants does
not rule out the possible existence of enough population~III stars to
solve the G-dwarf problem.  Since a metal-free early population should
contain no stars with masses of a solar mass or less, there would also
be no conflict with the lack of evidence for a non-standard IMF among
the stars in the Galactic halo and elsewhere.  The only apparent
constraint on, and evidence for, an early population of metal-free stars
would then be any distinctive abundance patterns that might be left
behind in their nucleosynthetic products.  Unusual abundance patterns
are indeed observed in some of the most metal-poor stars (McWilliam
1997), and some of them could result from enrichment by population~III
stars; for example, Norris, Ryan \& Beers (1997) suggest that the high
carbon-to-iron ratios that they find in many very metal-poor stars with
${\rm [Fe/H]} < -2.5$ could reflect enrichment by metal-free stars with
masses of about 20--30$\,$\Msun, which according to  Woosley \& Weaver
(1995) are predicted to yield high carbon-to-iron ratios.

    We conclude that population~III stars could play a role in solving
the G-dwarf problem in galaxies, and perhaps also in accounting for the
anomalous abundance ratios that have been seen in some of the most
metal-poor stars.  Population~III stars might also account for the
reionization of the universe that is needed to explain the highly
ionized state of the intergalactic medium (Haiman \& Loeb 1997). 
However, once again, it does not seem likely that large amounts of dark
matter can be accounted for by the remnants of such stars; this would
be possible only if large numbers of such stars were formed with high
efficiency at very high redshifts, which is not predicted by current
cosmological models.  Population~III stars may thus be most important
for their effects on the energetics and chemical enrichment of the early
universe.

\section{9~~CONCLUSIONS}

\noindent
    In the absence of any clear direct evidence for a time-varying IMF,
the possible existence of a top-heavy IMF at early times remains
hypothetical.  Some of the more extreme changes in the IMF that have
been proposed to explain various observations can probably now be ruled
out, but a number of lines of evidence continue to suggest at least
moderate variability of the IMF with time.  The longest-standing problem
in the theory of galactic chemical evolution, the G-dwarf problem,
persists and now even seems ubiquitous, suggesting that its solution may
be a general phenomenon such as a time-varying IMF, as was discussed in
Section~4.1.  The amount of variability in the IMF that is required to
solve the G-dwarf problem is moderate, amounting to an increase by only
a factor of $\sim\,$3 in the ratio of high-mass to low-mass stars formed
during the early stages of galactic evolution; this kind of variation is
not excluded by any of the constraints that have been discussed.

    The strongest evidence for variability of the IMF, with which most
authors concur, is the large total mass of heavy elements observed in
clusters of galaxies, mostly in the hot medium; the change in the IMF
needed to account for these observations is again moderate, typically
involving an increase by a factor of $\sim\,$2--3 in the ratio of
high-mass to low-mass stars formed during the early stages of cluster
evolution (Section~4.2).  The energetics of the hot gas may also provide
evidence that more supernovae were produced at early times than would be
predicted by a standard IMF (Section~4.3).  All of this evidence could
be accounted for if the slope $x$ of a power-law IMF were reduced by
about 0.4, or if the mass scale \m1\ in an IMF of form (1) or (2) were
increased by about a factor of 4 in young cluster galaxies.  The latter
type of variability is not implausible, as was discussed in Section~3,
and the steeper slopes of IMFs (1) and (2) at large masses yield better
agreement with the observed abundance ratios in clusters than does a
simple power law with a shallow slope at all masses (Section~4.2).

    Somewhat more direct evidence for time-variability of the IMF may
be provided by studies of high-redshift galaxies and by observations of
infrared background radiation that suggest high rates of massive star
formation at early times; these observations may require a top-heavy
early IMF in order not to predict too many low-mass stars that would
remain visible to the present time (Section~5).  The amount of variation
in the IMF that may be needed to account for these observations is again
of the same order as that required to account for the heavy-element
content of clusters of galaxies.  Thus, there appears to be a
significant body of evidence emerging that is consistent with the kind
of moderate variability of the IMF with time that has been discussed in
this paper.  Clearly, however, all of the evidence and constraints that
have been discussed here need to be strengthened before firm conclusions
can be drawn.

    Similar moderate variability of the IMF cannot, however, produce
large variations in the mass-to-light ratios of galaxies or account for
large amounts of dark matter in the form of stellar remnants; this
requires more extreme assumptions about the form of the early IMF, which
must be strongly truncated at the lower end (Section~6.).  Metal-free
`population~III' stars might have had such a strongly truncated lower
IMF, since the lower mass limit for such stars is predicted to be above
a solar mass (Section~7.).  Thus, massive population~III stars could
have existed in significant numbers without conflicting with the absence
of solar-mass metal-free stars, and they might help to account for some
of the evidence that has been attributed to a top-heavy early IMF\null. 
In principle they could also contribute significantly to cosmic dark
matter by the production of massive black-hole remnants, but even if
they do not, they can still contribute importantly to the energetics
and chemical enrichment of the early universe.  Further study of the
processes of early star formation and stellar evolution will thus be of
great importance for developing a better astrophysical understanding of
the early universe and of the origin of the stars and galaxies that we
now see.

\section{REFERENCES}

{\leftskip=5mm \parindent=-5mm

Adams F. C., Laughlin, G., 1996, ApJ, 468, 586

Alcock C., et al.\ (the MACHO collaboration), 1997, ApJ, 486, 697;
  see also ApJ, 490, L59 and ApJ, 491, L11

Angeletti L., Giannone, P., 1997, A\&A, 321, 343

Anninos P., Norman M. L., 1996, ApJ, 460, 556

Arimoto N., Yoshii Y., 1987, A\&A, 173, 23

Basri G., Marcy G. W., 1997, in Holt S. S., Mundy L. G., eds.,
  Star Formation Near and Far. AIP, Woodbury, NY, p.~228

Blain A. W., Smail I., Ivison R. J. \& Kneib J.-P., 1998, MNRAS, in
  press (astro-ph/9806062)

Bressan A., Chiosi C., Fagotto F., 1994, ApJS, 94, 63

Burkert A., Silk J., 1997, ApJ, 488, L55

Burstein D., Bender R., Faber, S. M., Nolthenius, R., 1997, AJ, 114,
  1365

Carr B., 1994, ARA\&A, 32, 531

Cen R., Ostriker J. P., Peebles P. J. E., 1993, ApJ, 415, 423

Chabrier G., M\'era D., 1997, A\&A, 328, 83

Chabrier G., Segretain L., M\'era D., 1996, ApJ, 468, L21

Dahn C. C., Liebert J., Harris H. C., Guetter H. H., 1995, in
  Tinney C. G., ed., The Bottom of the Main Sequence and Beyond.
  Springer-Verlag, Berlin, p.~239

David L. P., 1997, ApJ, 484, L11

David L. P., Jones C., Forman W., 1995, ApJ, 445, 578

David L. P., Jones C., Forman W., 1996, ApJ, 473, 692

Dwek E., Arendt R. G., Hauser M. G., Fixsen D., Kelsall T., Leisawitz D.,
  Pei Y. C., Wright E. L., Mather J. C., Moseley S. H., Odegard N.,
  Shafer R., Silverberg R. F., Weiland J. L., 1998, ApJ, in press
  (astro-ph/9806129)

Elbaz D., Arnaud M., Vangioni-Flam E., 1995, A\&A, 303, 345

Ellis R. S., 1997, ARA\&A, 35, 389

Elmegreen B. G., 1997, ApJ, 486, 944

Faber S. M., 1977, in Tinsley B. M., Larson R. B., eds., The Evolution
  of Galaxies and Stellar Populations.  Yale University Observatory,
  New Haven, p.~157

Ferguson H. C., Babul A., 1998, MNRAS, 296, 585

Festin L., 1998, A\&A, 333, 497

Fields B. D., Mathews G. J., Schramm D. N., 1997, ApJ, 483, 625

Flynn C., Gould A., Bahcall J. N., 1996, ApJ, 466, L55

Friel, E. D., 1995, ARA\&A, 33, 381

Gibson B. K., 1997, MNRAS, 290, 471

Gibson B. K., Matteucci F., 1997a, ApJ, 475, 47

Gibson B. K., Matteucci F., 1997b, MNRAS, 291, L8

Gibson B. K., Loewenstein M., Mushotzky R. F., 1997, MNRAS, 290, 623

Gnedin N. Y., Ostriker J. P., 1992, ApJ, 400, 1

Goodman A. A., Barranco J. A., Wilner D. J., Heyer M. H., 1998, ApJ,
  504, in press

Graff D. S., Laughlin G., Freese K., 1998, ApJ, 499, 7

Guiderdoni B., Bouchet F. R., Puget J.-L., Lagache G., Hivon E., 1997,
  Nature, 390, 257

Guzm\'an R., Lucey J. R., Bower R. G., 1993, MNRAS, 265, 731

Guzm\'an R., Gallego J., Koo D. C., Phillips A. C., Lowenthal J. D.,
  Faber S. M., Illingworth G. D., Vogt N. P., 1997, ApJ, 489, 559

Haiman Z., Loeb A., 1997, ApJ, 483, 21

Haiman Z., Thoul A. A., Loeb A., 1996, ApJ, 464, 523

Hayashi A., Eriguchi Y., Hashimoto M., 1998, ApJ, 492, 286

Hauser M. G., Arendt R. G., Kelsall T., Dwek E., Odegard N., Weiland J.
  L., Freundreich H. T., Reach W. T., Silverberg R. F., Moseley S. H.,
  Pei Y. C., Lubin P., Mather J. C., Shafer R. A., Smoot G. F., Weiss R.,
  Wilkinson D. T., Wright E. L., 1998, ApJ, in press (astro-ph/9806167)

Hillenbrand L. A., 1997, AJ, 113, 1733

Hunter D. A., Light R. M., Holtzman J. A., Lynds R., O'Neil E. J.,
  Grillmair C. J., 1997, ApJ, 478, 124

Ishimaru Y., Arimoto N., 1997, PASJ, 49, 1

J\o rgensen I., Franx M., Kjaergaard P., 1996, MNRAS, 280, 167

Kawaler S. D., 1996, ApJ, 467, L61

Kuijken K., Gilmore G., 1991, ApJ, 367, L9

Larson R. B., 1972, Nature, 236, 21 and Nature Phys.\ Sci., 236, 7

Larson R. B., 1974, MNRAS, 169, 229

Larson R. B., 1985, MNRAS, 214, 379

Larson R. B., 1986a, MNRAS, 218, 409 (L86)

Larson R. B., 1986b, in Norman C. A., Renzini A., Tosi M., eds.,
  Stellar Populations. Cambridge Univ.\ Press, Cambridge, p.~101  

Larson R. B., 1987, Comments Astrophys., 11, 273

Larson R. B., 1992a, MNRAS, 256, 641

Larson R. B., 1992b, in Tenorio-Tagle G., Prieto M., S\'anchez F.,
  eds., Star Formation in Stellar Systems.  Cambridge Univ. Press,
  Cambridge, p.~125

Larson R. B., 1995, MNRAS, 272, 213

Larson R. B., 1996, in Kunth D., Guiderdoni B., Heydari-Malayeri M.,
  Thuan T. X., eds., The Interplay Between Massive Star Formation,
  the ISM and Galaxy Evolution.  Editions Fronti\`eres, Gif sur
  Yvette, p.~3

Larson R. B., 1998, in Burkert A., McCaughrean M. J., eds., The Orion
  Complex Revisited.  ASP, San Francisco, in press

Larson R. B., Dinerstein H. L., 1975, PASP, 87, 911

Lilly S. J., Le F\`evre O., Hammer F., Crampton D., 1996, ApJ, 460, L1

Loewenstein M., Mushotzky R. F., 1996, ApJ, 466, 695

Low C., Lynden-Bell D., 1976, MNRAS, 176, 367

Madau P., Ferguson H. C., Dickinson M. E., Giavalisco M., Steidel
  C. C, Fruchter A., 1996, MNRAS, 283, 1388

Madau P., Pozzetti L., Dickinson M., 1998, ApJ, 498, 106

Maeder A., 1992, A\&A, 264, 105; erratum in A\&A, 268, 833

Mart\'\i n E. L., Zapatero Osorio M. R., Rebolo R., 1998, in Rebolo R.,
  Mart\'\i n E. L., Zapatero Osorio M. R., eds., Brown Dwarfs and
  Extrasolar Planets.  ASP, San Francisco, p.~507

Massey P., 1998, in Gilmore G., Parry I., Ryan S., eds., The Stellar
  Initial Mass Function.  Proceedings of the 38th Herstmonceux
  Conference, in press

Matteucci F., 1994, A\&A, 288, 57

Mayor M., Queloz D., Udry S., 1998, in Rebolo R., Mart\'\i n E. L.,
  Zapatero Osorio M. R., eds., Brown Dwarfs and Extrasolar Planets.
  ASP, San Francisco, p.~140

Mazzei P., De Zotti G., 1996, MNRAS, 279, 535

McWilliam A., 1997, ARA\&A, 35, 503

M\'endez R. A., Minniti D., De Marchi G., Baker A., Couch W. J., 1996,
  MNRAS, 283, 666

Meusinger H., 1989, Astron.\ Nachr., 310, 29 and 115

Miller G. E., Scalo J. M., 1979, ApJS, 41, 513

Mirabel I. F., 1989, in Tenorio-Tagle G., Moles M., Melnick J., eds.,
  IAU Colloq.\ 120, Structure and Dynamics of the Interstellar Medium.
  Springer-Verlag, Berlin, p.~396

Mushotzky R., Loewenstein M., Arnaud K. A., Tamura T., Fukazawa Y.,
  Matsushita K., Kikuchi K., Hatsukade I., 1996, ApJ, 466, 686

Nakamura F., Umemura M., 1998, preprint

Norris J. E., Ryan S. G., Beers T. C., 1997, ApJ, 488, 350

Oort J. H., 1970, A\&A, 7, 381

Pagel B. E. J., 1989, in Beckman J. E., Pagel B. E. J., eds.,
  Evolutionary Phenomena in Galaxies, Cambridge Univ.\ Press,
  Cambridge, p.~201

Peebles P. J. E., 1987, ApJ, 315, L73

Peebles P. J. E., 1993, Principles of Physical Cosmology, Princeton
  Univ.\ Press, Princeton, p.~655

Puget J.-L., Abergel A., Bernard J.-P., Boulanger F., Burton W. B.,
  D\'esert F.-X., Hartmann D., 1996, A\&A, 308, L5

Rees M. J., 1976, MNRAS, 176, 483

Reid I. N., Yan L., Majewski S., Thompson I., Smail I., 1996, AJ, 112,
  1472

Renzini A., 1997, ApJ, 488, 35

Richer H. B., Fahlman G. G., 1997, in Holt S. S., Mundy L. G., eds.,
  Star Formation Near and Far.  AIP, Woodbury, NY, p.~357

Rocha-Pinto H. J., Maciel W. J., 1996, MNRAS, 279, 447

Rowan-Robinson M., Mann R. G., Oliver S. J., Efstathiou A., Eaton N.,
  Goldschmidt P., Mobasher B., Serjeant S. B. G., Sumner T. J., Danese
  L., Elbaz D., Franceschini A., Egami E., Kontizas M., Lawrence A.,
  McMahon R., Norgaard-Nielsen H. U., Perez-Fournon I.,
  Gonzalez-Serrano J. I., 1997, MNRAS, 289, 490

Salpeter E. E., 1955, ApJ, 121, 161

Scalo J. M., 1986, Fundam.\ Cosmic Phys., 11, 1

Scalo J., 1998, in Gilmore G., Parry I., Ryan S., eds., The Stellar
  Initial Mass Function.  Proceedings of the 38th Herstmonceux
  Conference, in press (astro-ph 9712317)

Schlegel D. J., Finkbeiner D. P., Davis M., 1998, ApJ, 500, 525

Schmidt M., 1963, ApJ, 137, 758

Schwarzschild M., Spitzer L., 1953, Observatory, 73, 77

Silk J., 1977, ApJ, 214, 152

Simon M., 1997, ApJ, 482, L81

Sommer-Larsen J., 1991, MNRAS, 249, 368

Smail I., Ivison R. J., Blain A. W., 1997, ApJ, 490, L5

Tegmark M., Silk J., Rees M. J., Blanchard A., Abel T., Palla F.,
  1997, ApJ, 474, 1

Tinsley B. M., 1980, Fundam.\ Cosmic Phys., 5, 287

Totani T., Yoshii Y., Sato K., 1997, ApJ, 483, L75

Trentham N., 1994, Nature, 372, 157

Truran J. W., Cameron A. G. W., 1971, Astrophys.\ Space Sci., 14, 179

Uehara H., Susa H., Nishi R., Yamada M., Nakamura T., 1996, ApJ, 473,
  L95

van den Bergh S., 1962, AJ, 67, 486

Vazdekis A., Casuso E., Peletier R. F., Beckman J. E., 1996, ApJS, 106,
  307

von Hippel T., Gilmore G., Tanvir N., Robinson D., Jones D. H. P.,
  1996, AJ, 112, 192

Wakker B. P., van Woerden H., 1997, ARA\&A, 35, 217

Woosley S. E., Weaver T. A., 1995, ApJS, 101, 181

Worthey G., Dorman B., Jones L. A., 1996, AJ, 112, 948

Worthey G., Faber S. M., Gonz\'alez J. J., 1992, ApJ, 398, 69

Wyse R. F. G., 1997, ApJ, 490, L69

Wyse R. F. G., Gilmore G., 1995, AJ, 110, 2771

Zepf S. E., Silk J., 1996, ApJ, 466, 114 (ZS)

}
\bye